\documentstyle[aps,prb,preprint,tighten,eqsecnum]{revtex}
\begin{document}
\title
{Quantum phases of the Shraiman-Siggia model}
\author{\large Subir Sachdev}
\address{Departments of Physics and Applied Physics, P.O. Box 6666,\\ 
Yale University, New Haven, CT 06511}
\date{\today}         
\maketitle
\begin{abstract}
We examine phases of 
the Shraiman-Siggia model of lightly-doped, square lattice
quantum antiferromagnets in a self-consistent,
two-loop, interacting magnon analysis. 
We find magnetically-ordered and quantum-disordered
phases both with and without incommensurate spin correlations.
The quantum disordered phases have a pseudo-gap in the 
spin excitation spectrum. The quantum transition between 
the magnetically ordered and commensurate
quantum-disordered  phases is argued to have
the dynamic critical exponent $z=1$ and the same leading critical behavior
as the disordering transition in the pure $O(3)$ sigma model.
The relationship to experiments on the doped cuprates is discussed. 
\end{abstract}
\narrowtext

\section{INTRODUCTION}
\label{intro}

The last few years have seen numerous experiments examining the magnetic
properties of the doped copper-oxide compounds in some detail. 
However, our theoretical understanding has not kept in step, partly due
to the numerous competing effects and ensuing complexity of these
materials: at low temperatures the effects of disorder are paramount,
and at larger doping there is the onset of superconductivity. 

On the theoretical side, a popular model for investigating the interplay
of doping and antiferromagnetic spin correlations has been the $t$-$J$ model.
Numerical and high temperature series studies of this model have been
especially valuable as a testing ground for various theoretical ideas.
In an important advance, Shraiman and Siggia~\cite{boris} (SS) proposed a
phenomenological description of the  
long-wavelength interplay between
spin and charge transport in this model. There are numerous 
theoretical reasons
for believing that their long-wavelength model is a correct description of the
$t$-$J$ model at low temperatures.
It is expected that the SS model is quite
robust, especially at low doping concentrations, 
and will describe a whole class of doped antiferromagnets, 
not just the $t$-$J$ model. 
However, at sufficiently large doping, an approach based upon
long-wavelength distortions of the background antiferromagnetic order
must eventually become invalid; the mechanism of this break-down of the SS model 
is not understood and remains an important open problem.
This paper shall examine the SS model
in a new approach which is designed to explicate the nature of 
the quantum-disordered phases {\em i.e.\/} the phases with no long-range
magnetic order. 
These phases appear at a reasonably small doping concentration, where 
it is expected, though not certain, that the mapping
between the SS and $t$-$J$ models is still valid.
We will also discuss the relationship of our results
to other theoretical work and some experiments in Section~\ref{conc}.

We begin by writing down a simplified version of the action $S$ of the
Shraiman-Siggia model:
\begin{equation}
S = \int_0^{\beta \hbar} d \tau \int d^2 r \left[ S_n + S_f + S_c \right],
\end{equation}
with $\tau$ the Matsubara time, $r = (x,y)$ the spatial co-ordinates,
$\beta = 1/(k_B T)$, and $T$ the absolute temperature,
The first part, $S_n$ describes fluctuations of the antiferromagnetic order
parameter $n_{\ell}$. Here $n_{\ell}$ is a 3-component vector and is
taken to have unit length 
\begin{equation}
\sum_{\ell=1}^{3} n_{\ell}^2 = 1.
\label{const}
\end{equation}
We then have
\begin{equation}
S_n = \frac{\rho_s^0}{2\hbar} \left( (\partial_x n_{\ell} )^2 +
(\partial_y n_{\ell} )^2 + \frac{1}{c_0^2}
\left(\frac{\partial n_{\ell}}{\partial \tau} \right)^2 \right)
\end{equation}
with $\rho_s^0$ the bare spin stiffness and $c_0$ the bare spin-wave
velocity. The momentum
of the $n_{\ell}$ field is restricted to be smaller than an ultraviolet
cutoff $\Lambda$, which is also the scale at which  the coupling constant
are defined. 

The action of the fermionic dopant holes is given by $S_f$. The holes are
describes by fermionic spinor fields $\Phi_{v\alpha}$ where $\alpha=\uparrow,
\downarrow$ is the spin index, and $v$ is `valley' index. The valleys
are regions in the Brillouin zone around the minima of the fermion dispersion
spectrum. For the square-lattice $t$-$J$ model there is a great deal of 
evidence that there are two vallleys centered at the points
$(\pi/2 , \pm \pi/2)$ in the reduced Brillouin zone~\cite{veit,variational}. 
We will rotate
our co-ordinate system from the conventional one, so that our $x$-axis
is at an angle of $45$ degrees to the axes of the square lattice,
and the principal axes of the
valleys are along the new $x$ and $y$ axes; the index $v$ will
therefore take the values $v=x,y$. Further, we will
measure momenta from the center of the valleys. With these conventions, the
fermionic action $S_f$ is 
\begin{eqnarray}
S_f =  && 
 \Phi_{x\alpha}^{\dagger}
\left( \frac{\partial}{\partial\tau} - \frac{\hbar^2}{2m_l} \frac{\partial^2}{
\partial x^2} - \frac{\hbar^2}{2 m_h} \frac{\partial^2}{\partial y^2}
\right) \Phi_{x\alpha}  \nonumber \\ 
&&~~~~  + \Phi_{y\alpha}^{\dagger}
\left( \frac{\partial}{\partial\tau} - \frac{\hbar^2}{2m_h} \frac{\partial^2}{
\partial x^2} - \frac{\hbar^2}{2 m_l} \frac{\partial^2}{\partial y^2}
\right) \Phi_{y\alpha},
\end{eqnarray}
where $m_l$, $m_h$ are the light and heavy masses of the hole. 

It is important to realize that there is no simple relationship between the
bare fermionic field $c_{i\alpha}^{\dagger}$ 
of the $t$-$J$ model ($i$ is the site index)
and the continuum fields $(n_{\ell}, \Phi_{v \alpha} )$ of the hydrodynamic
SS model. Crudely speaking, one may consider this field transmutation as
a form of spin-charge separation in which the spin-1/2, charge $e$
$c_{i\alpha}^{\dagger}$ fermionic field 
has separated into the spin-1, charge 0,
bosonic
$n_{\ell}$ quanta and the spin-1/2, charge $e$, $\Phi_{v\alpha}$ fermions. 

Finally, we will consider only the leading term in the coupling, $S_c$ 
between the holes and the $n_{\ell}$ field, which is
 responsible for inducing local incommensurate spin correlations.
It is expected that the remaining terms in the SS model 
are innocuous and do not 
change the results of this paper qualitatively. We have
\begin{equation}
S_c =  \kappa  \left( \Phi_{v\alpha}^{\dagger}
\sigma^{\ell}_{\alpha\beta} \Phi_{v\beta} \right) 
\left(\epsilon_{\ell m p} n_{m} \partial_v n_{p} \right)
\label{sc}
\end{equation}
where $\sigma^{\ell}$ are three Pauli matrices, and $\kappa$ is a coupling
constant. A crucial feature of the SS model is that there is no
three-body coupling like $n_{\ell} \Phi_{v\alpha}^{\dagger} 
\sigma^{\ell}_{\alpha\beta} \Phi_{v\beta}$ between the fermions
and $n_{\ell}$ quanta~\cite{comment}: 
such a term is forbidden by a sublattice interchange
symmetry of the SS model~\cite{boris} under which $n_{\ell}$ change sign
while the $\Phi_{v\alpha}$ remain invariant. Other models of nearly
antiferromagnetic fermi liquids~\cite{hertz,Andy} posses such a term
and as a result, have quantum disordered phases with rather different 
properties.
 
We wish to clarify a crucial point about our particular form of $S$
at the outset. SS have shown that there is
a mapping, in principle exact, to an alternative form of $S$
in which the degrees of freedom are a spin-1/2, charge 0, complex scalar
$z_{\alpha}$ with $\alpha = \uparrow,\downarrow$ and {\em spinless},
charge $e$
fermions $\Psi_{va}$
where $a=A,B$ is a sublattice index. The $(n_{\ell}, \Phi_{v\alpha})$
fields are related to the $(z_{\alpha}, \Psi_{va})$ fields by
\begin{equation}
n_{\ell} = z_{\alpha}^{\ast} \sigma^{\ell}_{\alpha\beta} z_{\beta}
~~;~~ \left( \begin{array}{c} \Phi_{v\uparrow} \\ \Phi_{v\downarrow}
\end{array} \right) = \left( \begin{array}{cc} z_{\uparrow}, & 
-z_{\downarrow}^{\ast} \\ z_{\downarrow} & z_{\uparrow}^{\ast} \end{array}
\right) \left( \begin{array}{c} \Psi_{vA} \\ \Psi_{vB} \end{array} \right)
\end{equation}
Note that there is a $U(1)$ gauge transformation on the 
$(z_{\alpha}, \Psi_{va})$ fields which leaves the 
$(n_{\ell}, \Phi_{v\alpha})$ fields invariant:
\begin{equation}
z_{\alpha} \rightarrow z_{\alpha} e^{i \phi} ~~~;~~~\Psi_{vA}
\rightarrow \Psi_{vA} e^{-i\phi}~~~;~~~\Psi_{vB} e^{i \phi},
\end{equation}
where $\phi$ has an arbitrary dependence on spacetime. The action
in terms of the $(z_{\alpha}, \Psi_{va})$ will therefore also have to
be invariant under the gauge transformation; there is no such restriction
on the action in the $(n_{\ell}, \Phi_{v\alpha})$ variables.
This latter absence of gauge restrictions, and associated 
long-range gauge forces will be quite useful to us in our analysis.

SS go on to state that equivalent results are obtained in computations
using either choice of fields. In principle, this statement is correct. 
In practice, however, one is usually 
restricted in the analysis to perturbation theory, in which the 
quantum numbers of the low-lying excitations are essentially identical
to those of the degrees of freedom in the action. 
Equivalent results in both theories have been obtained
in the vicinity of magnetically ordered states, where one is performing
a small fluctuation, spin-wave analysis. On the other hand, 
the choice of fields is expected to have dramatic consequences in a 
quantum-disordered state.
Kane {\em et. al.\/} and others~\cite{kane}
 made a choice of 
fields equivalent to
the $(z_{\alpha} , \Psi_{va})$ formulation of the SS model. 
Thus, not surprisingly,
their Schwinger-boson mean-field theory 
yielded a quantum disordered phase with
deconfined, spin-1/2, charge 0, bosonic spinons (the $z_{\alpha}$) 
and spinless, charge $e$ fermions
($\Psi_{va}$).
 
In this paper, we use another approach and 
shall examine the phases of the SS model which are
obtained naturally in the $(n_{\ell} , \Phi_{v\alpha})$ formulation.
Some of the results of our calculations were noted some time 
ago~\cite{jinwu}.
While, in the end, we have no formal justification for claiming that
the $(n_{\ell}, \Phi_{v\alpha})$ formulation is more accurate
than the $(z_{\alpha}, \Psi_{va})$ approach, we can offer the following
motivations.
The
$n_{\ell}$ field formulation has been quite successful in describing
the undoped, frustrated antiferromagnet.
The long wavelength action of the undoped antiferromagnet~\cite{CHN} 
is simply the
$O(3)$ sigma model, $S_n$, and it is expected to display a quantum-disordered
phase in which massive $n$ quanta form the lowest excitations and 
carry spin 1~\cite{polyakov}.
Consistent with this, 
recent investigations of the quantum-critical behavior in these systems
have argued for the superiority of the $n$-field based approach,
and have successfully explained a number of experimental and numerical
computations on the square lattice antiferromagnet~\cite{jinwu,CSY}. Further,
careful analysis of fluctuations~\cite{Sach-Read1} in 
the $z_{\alpha}$-based theories of the undoped
antiferromagnet has yielded quantum disordered phases in which the quantum
numbers of the low-lying excitations are identical to those $n$-field,
which means that fluctuations in fact confine
bosonic spinons into $S=1$ particles.
It is then natural to explore the consequences of doping
in a model in which the correct physics in the limit of zero doping is
captured most directly {\em i.e.\/} in the $(n_{\ell}, \Phi_{v\alpha})$
approach. We shall argue later in this paper, that the results of such
an investigation are consistent with the available numerical and 
experimental data on doped antiferromagnets. 

\subsection{Summary of Results}
We now discuss the main results of our calculations.
We will distinguish the quantum phases by the properties of the 
equal-time $n_{\ell}$ correlator  
\begin{equation}
\tilde{S} (q) = \langle | n_{\ell} ( q) |^2 \rangle
\end{equation}
as $T \rightarrow 0$, as a function of the momentum $q$.
We emphasize that $\tilde{S} (q)$ is {\em not\/} the full structure
factor $S(q)$ measured in neutron scattering experiments; $S(q)$ will
contain additional terms involving the contribution of the 
spin-1/2 $\Phi_{v\alpha}$ fields. Phases with magnetic long-range order
have a delta function term in $\tilde{S} (q)$ at $T=0$; this delta function
is at $q=0$ in the commensurate long-range-ordered phase (hereafter
referred to as CLRO) and at $q\neq 0$ for the case of 
incommensurate long-range-order (ILRO) ($q$ is measured from ($\pi,\pi$)). 
The quantum disordered phases
have no delta-function terms at $T=0$,
but only a peak of finite, though possibly
small, width. This peak is at $q=0$ in the commensurate quantum-disordered
phase (CQD) and at $q\neq 0$ in the incommensurate case (IQD).
In agreement with the SS analysis~\cite{boris} and 
experimental results~\cite{aeppli}, the peak in the
ILRO and IQD phases was found to occur along the 
conventional $(1,0)$ and $(0,1)$
axes of the square lattice (these are the $(1,1)$ and $(1,-1)$ axes
in our rotated co-ordinate system).

We will have little to add here to existing 
studies~\cite{boris,frenkel,borislast,gorkov} of the
properties of the magnetically-ordered phases (CLRO and IRLO): their low-lying
excitations are spin-waves involving long-wavelength deformations of the
ordered state. Our focus will mainly be upon the new quantum-disordered
phases (CQD and IQD) and their unusual properties. The $n_{\ell}$
quanta in both phases were found to be fully gapped. The low-lying excitations
in the $n_{\ell}$ sector consist of a triply-degenerate spin-1 particle
with a finite energy. However, the spin-1/2 $\Phi_{v\alpha}$
particles continue to form a Fermi sea which possesses gapless
fermionic excitations with charge $e$ and spin-1/2. Despite the presence
of these gapless excitations, the $n_{\ell}$ gap is robust as there
is no term in the SS model which permits the decay of a $n_{\ell}$
quantum to a fermion particle-hole pair. The importance of the absence
of the three-body term noted above, is now evident. Taken as a whole,
the model thus only has a pseudo-gap to spin excitations in the
CQD and IQD phases. One of the consequences of the presence of the
gapless spin-1/2 fermions is that the uniform spin susceptibility
of the CQD and IQD phases will be finite at $T=0$ due to the Pauli 
contribution~\cite{comment}.
We also note, that our calculation has completely neglected the effect
of Berry phases; in the context of undoped antiferromagnets
it has been argued~\cite{Sach-Read1} 
that Berry phases should induce spin-Peierls 
long-range order in the CQD phases. It is possible that such spin-Peierls
order will also exist in the CQD phase of the doped antiferromagnet.

We undertook a partial numerical survey of the phase diagram of
the SS model as a function of $\rho_s^0$, $\kappa$ and the hole density.
Parameters were always chosen so that the zero doping state
was CLRO. This CLRO state was always found to be stable over a small,
but finite, doping concentration.
Over some of the regime examined, the sequence of phases with
increasing doping was CLRO - CQD - IQD. We studied the $T=0$ quantum
transition between the CLRO and CQD phases and will present 
evidence indicating
that it has dynamic critical exponent $z=1$ and the same leading
critical behavior as the transition in the pure $O(3)$ sigma model;
however, the corrections to scaling in the two models were found to be
quite different. The boundary between the CQD and IQD phases is an
example of a disorder line~\cite{disorder}: 
our calculation only found a non-analyticity
in the dependence of the structure factor on the bare coupling constants
at the disorder line,
but no strong long-wavelength fluctuations.

In a region of the phase diagram with 
$\kappa$ large, we found the sequence
CLRO - ILRO with increasing doping. In principle, there should eventually
be a ILRO to IQD transition, but for the parameters examined, 
we did not find one before a 
doping level where the incommensuration wavevector was almost as large
as the momentum upper cutoff. 

We will argue from the above numerical results, and from theoretical 
considerations, that there is a Lifshitz point in the $\rho_s^0$, $\kappa$
plane where all the four phases - CLRO, ILRO, CQD, IQD - meet. 
Some properties of this multicritical point will be discussed. 
  
We also have extensive results on the temperature dependence of
equal-time correlation functions in the various phases. In particular,
the temperature dependences in the spin correlation length and
the structure factor are quite instructive, and will be described later.

\section{CALCULATIONS}
\label{calcs}

For the case of the undoped antiferromagnet, the $1/N$ expansion
on the $O(N)$ non-linear sigma model offers a convenient and accurate
method for exploring properties in the vicinity of $T=0$ quantum 
transitions~\cite{CSY}.
The extension of the $1/N$ expansion to the doped antiferromagnet is
however not straightforward because of the presence of
the third-rank 
$\epsilon_{\ell m p}$ tensor in $S_c$ (Eqn. (\ref{sc})), which is
special to the case $N=3$. Even with this complication, it is
still possible to justify perturbative 1/N calculations, although
in a rather
inelegant way: after the fermions have been integrated out, the coupling
$\kappa$  in  the effective action of the 
$n$ field has be to scaled by $1/N^{\mu}$ where $0 < \mu < 1/2$.
Not much is learned from this extension to general $N$, and we will
therefore spare the reader the details. 
We will be satisfied, instead, in restricting our discussion to the special
case of $N=3$, and viewing our $1/N$ calculation as a physically motivated,
self-consistent, interacting magnon approximation. 
The magnon-magnon interactions
are computed in a manner which is directly inspired by the $1/N$ expansion
of the undoped antiferromagnet~\cite{CSY}.

An important property of our approach is that spin-rotation invariance is
explicitly preserved at all stages of the calculation. This is crucial
for a proper study of the quantum disordered phases of the model, especially
when the $n$ quanta acquire a gap. We thus expect our approximations
to work best in the quantum-disordered phase, at the $T=0$ quantum
transition, and in the intermediate temperature 
quantum-critical region~\cite{CHN,CSY}.
At the same time, the low temperature properties
 in a region with magnetic long-range
order (the renormalized classical region~\cite{CHN})
in the ground state, may not be well described.
Even in the undoped antiferromagnet, the $1/N$ expansion is singular in the
renormalized-classical region, and a careful interpretation of the
results is required~\cite{CSY}. We therefore will focus below 
here will be mainly on the quantum-disordered phases phases.

We begin by expressing the action in suitable dimensionless parameters.
We rescale lengths such that the upper-cutoff in momentum space for the
$n$ field is 1; thus 
\begin{equation}
r \rightarrow \frac{r}{\Lambda}
\end{equation}
Similarly, the times $\tau$ are rescaled so that the bare spin-wave velocity
of the $n$ field is unity:
\begin{equation}
\tau \rightarrow \frac{\tau}{c_0 \Lambda}
\end{equation}
After rescaling the temperature
\begin{equation}
T \rightarrow T \frac{\hbar c_0 \Lambda }{k_B}
\end{equation}
we then have the following modified form of $S_n$
\begin{equation}
S_n = \frac{1}{2g} \left( (\partial_{x} n_{\ell} )^2 +
(\partial_{y} n_{\ell} )^2 + (\partial_{\tau} n_{\ell})^2 \right)
\end{equation}
where the dimensionless coupling constant $g$ is given by
\begin{equation}
g = \frac{\hbar c_0 \Lambda}{\rho_s^0}
\end{equation}
The fermionic action will retain its form after rescaling the field
$\Phi \rightarrow \Lambda \Phi$ and rescaling to the dimensionless
effective masses
\begin{equation}
m_{h,l} \rightarrow m_{h,l} \frac{\Lambda \hbar}{c_0}
\end{equation}
Finally in $S_c$ we replace $\kappa \rightarrow \kappa c_0$.

We will impose the constraint (\ref{const}) by a Lagrange multiplier field
$\lambda$. Thus we need to evaluate the functional integral
\begin{eqnarray}
Z = \int {\cal D} n_{\ell} {\cal D} \Phi_{v\alpha} && {\cal D} \lambda 
\exp \Biggl(
- S  \nonumber \\ 
 && - i \left. \int_0^{1/T} d\tau \int d^2 x \frac{\lambda}{2g} ( n_{\ell}^2 - 1)
\right)
\end{eqnarray}

It is now possible to set up a rotationally-invariant, diagrammatic
expansion of all observables associated with $Z$. 
We will work at finite $T$, and so no breaking of spin rotation invariance
can occur; the properties of the ground state will be elucidated by taking
the $T \rightarrow 0$ limit.
We treat the $\lambda$
field in much the same way as in the undoped system~\cite{polyakov}.
We assume that fluctuations of $i\lambda$ occur about a saddle point value
$\overline{\lambda}$; we therefore write
\begin{equation}
i \lambda = \overline{\lambda} + i \tilde{\lambda}
\end{equation}
where $\tilde{\lambda}$ is the fluctuating part of $\lambda$.
The value of $\overline{\lambda}$ is to be determined at the
end of the calculation to satisfy the constraint (\ref{const}).
The diagrammatic expansion now has three bare propagators: the conventional
Green's function of the fermions $\Phi_{va}$, the propagator, $G^0$,
of the $n_{\ell}$ field
\begin{equation}
G^0 ( q, i\omega_n ) = \frac{1}{q^2 + \omega_n^2 + \overline{\lambda}},
\label{g0}
\end{equation}
($q$ is the wavevector and $\omega_n$ is a Matsubara frequency)
and the propagator $1/\Pi$ of $\tilde{\lambda}$, with~\cite{polyakov}
\begin{equation}
\Pi ( q, i\omega_n) = 
T \sum_{\epsilon_n} \int \frac{d^2 k }{4 \pi^2} 
G^0 ( k+q, i\epsilon_n+i\omega_n ) G^0 ( k , i\epsilon_n ).
\label{Pi}
\end{equation}
There are two interaction vertices: the four-body $(n-n-\Phi-\Phi)$
coupling in $S_c$ and a three-body $(\tilde{\lambda}-n-n)$ vertex with the value
$i/(2 g)$. Finally, there is a rule to prevent over-counting:
no $\tilde{\lambda}$ propagator can be followed by a bubble consisting
just of two $G^0$ propagators.

We may now write the fully-renormalized correlator of the $n$ field in the
form
\begin{equation}
G ( q, i\omega_n ) = \frac{1}{q^2 + \omega_n^2 + m^2 + 
\Sigma ( q, i\omega_n ) - \Sigma (0, 0)}
\label{Gfunc}
\end{equation}
where $\Sigma$ is the self energy and the `mass' $m$ is given by
\begin{equation}
m^2 = \overline{\lambda} + \Sigma (0, 0)
\end{equation}
The lowest order contributions to $\Sigma$ from magnon-magnon ($\Sigma_n$)
and magnon-fermion ($\Sigma_f$) interactions are shown in Fig~\ref{feyn}, 
and their values are
\widetext
\begin{eqnarray}
\Sigma &=& \Sigma_n + \Sigma_f \nonumber \\
\Sigma_n ( q, i\omega_n ) &=& \frac{2}{3} T \sum_{\epsilon_n}
\int \frac{d^2 k }{4 \pi^2}  
\frac{G^0 ( k+q , i\epsilon_n + i\omega_n )}{\Pi (k , i\epsilon_n )} \nonumber \\
\Sigma_f ( q, i\omega_n ) &=& -4 g^2 \kappa^2 T \sum_{\epsilon_n}
\int \frac{d^2 k }{4 \pi^2} (2 q_v + k_v )^2 \chi_v ( k , i\epsilon_n ) 
G^0 ( k+q , i\epsilon_n + i\omega_n )
\label{Sigma}
\end{eqnarray}
where $\chi_v$ is polarization of the fermion in valley $v$. We used
the following expression, appropriate for an elliptical Fermi surface:
\begin{equation}
\chi_x ( q, i\epsilon_n ) = \chi_0 (T) \left( 1 - \frac{|\epsilon_n |}{
\left[ \epsilon_n^2 + v_F^2 \left( q_x^2 (m_h/m_l)^{1/2}
+ q_y^2 (m_l / m_h )^{1/2} \right) \right]^{1/2}} \right)
\end{equation}
\narrowtext
and similarly for $\chi_y$. The Fermi velocity $v_F$, the Fermi wavevector
$k_F$ and the doping concentration $\delta$ are related by the equations
\begin{equation}
v_F = \frac{k_F}{\sqrt{m_l m_h}}~~~~~~~;~~~~~~~~\delta = \frac{k_F^2}{\pi}.
\end{equation}
and the polarization is taken to vanish unless
\begin{equation}
\left( q_x^2 (m_h/m_l)^{1/2}
+ q_y^2 (m_l / m_h )^{1/2}\right)^{1/2} < 2 k_F
\end{equation}
The prefactor $\chi_0$ we chose as the compressibility of a free Fermi gas
at a temperature $T$:
\begin{equation}
\chi_0 (T) = \frac{\sqrt{m_l m_h}}{2 \pi} \left( 1 - e^{-k_F v_F / 
(2 T)} \right) 
\label{chival}
\end{equation}

Our approach consisted of using the above approximation for $\Sigma$
and then solving the constraint equation (\ref{const}), or
\begin{equation}
3 g T \sum_{\omega_n} \int \frac{d^2 q}{4 \pi^2} G (q, i\omega_n) = 1
\label{constg}
\end{equation}
for the value of $m^2$ (or equivalently $\overline{\lambda}$).
The dependence of $\Sigma$ on $G^{0}$ was made partially self
consistent by replacing $\overline{\lambda}$ by $m^2$ in (\ref{g0}),
thus using $G^0 (q, i\omega_n )= 1/(q^2 + \omega_n^2 + m^2)$
in (\ref{Pi}) and (\ref{Sigma}). A fully self-consistent approach would
require we replace $G^0$ by $G$ in these equations;
this is computationally much more difficult and was not numerically
implemented.
Our approximation thus amounts to replacing $G^0$ by $G$, but
then ignoring the momentum and frequency 
dependence of the self energy in $G$.
For the most part, this omission is not expected to be serious, as
corrections can be organized order by order in $\kappa$. However,
we cannot rule out the possibility, especially in the
quantum-disordered phases, that there is some entirely
different, possibly gapless, solution of the fully self-consistent
equations; such a solution will clearly be non-perturbative in $\kappa$.
We also note here that in our analytical considerations below of the 
boundaries between the phases, we will
include the full $G$ in (\ref{Sigma}).

The numerical determination $m^2$ as a function of $T, \delta$ was
carried out on a HP-RISC workstation. A meaningful solution always
existed at all finite $T$, with no phase transitions as a function
of $T$ or $\delta$. Phase transitions are however present at $T=0$,
and were examined by studying the $T \rightarrow 0$ limit of our solutions.
The computations required about 3 weeks of computer time.

\section{RESULTS}

We now describe the results of our numerical calculations.
The nature of the ground state can be determined from the values
and $T$ dependences of $q_c$ and $\overline{m}$ where
$q=q_c$ is the location of the maximum of $G(q, i\omega_n = 0)$,
and
\begin{equation}
\overline{m}^2 = m^2 + \Sigma (q_c , i\omega_n = 0) - \Sigma(0,0)
\label{defmb}
\end{equation}
It is easy to see from (\ref{Gfunc}) that $\overline{m}$ 
is roughly the inverse correlation
length (`roughly' because this
neglects $\partial \Sigma/ \partial q^2$;
including this term yields  corrections of the order of unity which are not
strongly $T$ dependent).
The various phases can be identified by studying the $T$ dependence of
$\overline{m}$ as $T \rightarrow 0$, as will be described below.
The values of $q_c$ were approximately $T$ independent
and distinguish between commensurate and incommensurate phases.

Two samples of our results are contained in Figs.~\ref{mbar1} and~\ref{mbar2}
which plot the $T$ and doping dependence of $\overline{m}$ for
two sets of coupling constants. For completeness we also show 
in Figs~\ref{m1} and~\ref{m2} the values of $m$ for the same samples.
We will now describe the properties of the phases in these
figures and follow that up with some general discussion
on the nature of the quantum transitions between them.
 
\subsection{Long-range-ordered states}
The states with magnetic order are expected
to have $\overline{m} \rightarrow 0$ as $T \rightarrow 0$.
In particular, the low $T$ dependence should be~\cite{brezin}
\begin{equation}
\overline{m} \sim \exp\left( - \frac{2 \pi \rho_s}{k_B T} \right)
\end{equation}
where $\rho_s$ is the fully renormalized spin stiffness.  
Numerical solutions at very lower $T$ took longer times to converge,
so it was difficult to see this exponential behavior in some of the
doped samples. We simply identified the samples in which
$\overline{m}$ vanished with an upward curvature as $T \rightarrow
0$, as possessing magnetic long-range order. Further states with
$q_c = 0$ ($q_c \neq 0$) were identified as CLRO (ILRO).

\subsection{Quantum disordered states}
These states have $\overline{m}$ saturating at a finite value as 
$T \rightarrow 0$, which is roughly the gap, $\Delta \sim \overline{m}
(T \rightarrow 0)$, in the $n$ sector.
 Again, the value of $q_c$ distinguishes between
the CQD and IQD states.

We examine the nature of the spin correlations at a point 
in the IQD phase by
plotting the 
$n$ field contribution to the
structure factor, $\tilde{S}(q)$,
\begin{equation}
\tilde{S}(q) = g T \sum_{\omega_n} G(q, \omega_n)
\end{equation}
in Fig.~\ref{sfac}. Notice that there is strong overlap between the
peaks at high temperature. Upon lowering the temperature, the peaks
first sharpen considerably, but then their width saturates.

Let us discuss the form of the $n$ spectrum at $T=0$ in the CQD phase.
We will focus on real frequencies, $\omega$ just above the gap $\Delta$,
and small momenta $q$. The magnon contribution to the self energy, $\Sigma_n$
in (\ref{Sigma})
does not acquire an imaginary part until $\omega = 3 \Delta$ and can
therefore be completely ignored. The damping from the fermion particle-hole
continuum, $\Sigma_f$ is however not so innocuous. We find~\cite{square}
\begin{equation}
\mbox{Im} \Sigma_f ( q, \omega ) \sim \left\{ 
\begin{array}{ll}
0         &      ~~ 0< \omega < \Delta \\
|q| (\omega - \Delta)^2  & 
~~0 < \omega - \Delta \ll {\displaystyle \frac{q^2}{2 \Delta}} \\
 \Delta^{1/2} (\omega - \Delta)^{5/2}  & 
~~\omega - \Delta \gg {\displaystyle \frac{q^2}{2 \Delta}} 
\end{array}
\right.
\end{equation}
and $\mbox{Im} \Sigma_f ( q, - \omega ) = - \mbox{Im} \Sigma_f ( q,\omega )$.
The $n$ spectral weight is then given by
\begin{equation}
G ( q, \omega ) \sim \frac{1}{\Delta^2 + q^2 - \omega^2 + \Sigma_f (q, \omega )}
\end{equation}
From the above results it follows that at $q=0$ we have
\begin{equation}
\mbox{Im} G ( q=0, \omega \geq \Delta ) 
= \frac{a_1}{\Delta} \delta ( \omega - \Delta ) + \frac{a_2}{ 
\Delta^{3/2}} ( \omega - \Delta)^{1/2}
\end{equation}
for some constants $a_1 , a_2$. Thus there is a sharp spin-1 quasiparticle peak,
and a 
second background term which is a
direct consequence of the coupling of the $n$ quanta to the
particle-hole continuum.
At small, but finite $q$, the sharp peak moves to 
$\omega \sim \Delta + q^2 /(2 \Delta)$ and acquires a finite width; there
is absorption at all frequencies greater than $\Delta$. 

The spectral properties of the IQD phase are essentially identical except that
the role of the point $q=0$ is replaced by $q=q_c$; in obtaining this
result it is, of course, necessary to replace $G^0$ by $G$ in (\ref{Sigma}).

\subsection{Quantum transitions}

Our results in Figs.~\ref{mbar1} and~\ref{mbar2} show two sequences of
quantum transitions with increasing doping: CLRO-CQD-IQD and CLRO-ILRO.
In the second case there should eventually
be a ILRO to IQD transition, but for the parameters examined, 
we did not find one before a 
doping level where the incommensuration wavevector was almost as large
as the momentum upper cutoff. 
We now present a theoretical analysis of some issues raised by the
existence of these
quantum transitions.

\subsubsection{CLRO to CQD quantum transition}
An important ingredient in determining
the universality class of this transition is the analytic structure of
the $n$ field self-energy $\Sigma_f$ as $T\rightarrow 0$ in the N\'{e}el
phase and the quantum-critical point. 
We consider (\ref{Sigma}) in the limit $m\rightarrow 0$,
and $T\rightarrow 0$ when the frequencies become continuous variables,
and the Matsubara summations can be converted to integrations.
It is evident that $\Sigma_f ( q, i\omega )$ is an even function
of $\omega$. Moreover, it is not difficult to show that there are no
infra-red divergences in either
$\partial \Sigma_f / \partial q^2 |_{q=\omega=0}$ or
$\partial \Sigma_f / \partial \omega^2 |_{q=0, \omega \searrow 0}$.
This implies that for $q, \omega$ small we have
\begin{equation}
\Sigma_f ( q, i\omega ) = \Sigma_f (0,0) + b_1 q^2 + b_2 \omega^2 + \ldots
\label{sigexpand}
\end{equation}
Thus the gapless fermion particle-hole sea 
has not induced any non-analyticities
in $\Sigma_f$ to this order. There are indeed non-analytic terms
present  at higher order in
$\Sigma_f$ which are signaled by infra-red divergences in higher derivatives
of $\Sigma_f$; we will discuss the form of such terms below. 
For our purposes, it is sufficient to note here that all such higher gradient
terms are expected to be irrelevant
at the CLRO to CQD transition. Thus the gapless fermion particle-hole 
excitations have had a relatively innocuous effect: they have mainly
lead to renormalizations of the spin-wave velocity and spin
stiffness. The universality class of the CLRO-CQD transition is thus
expected to be the same as that in the undoped sigma model. 
This is a transition with dynamic critical exponent $z=1$ and its
leading universal properties have been discussed in some detail 
by Chubukov {\em et. al.\/}~\cite{CSY}. 
All of the scaling functions of Chubukov {\em et. al.\/}~\cite{CSY} should
therefore also apply to the present doped antiferromagnet. The main
effect of the fermions has been to change the value of the effective
coupling constant and renormalize the spin-wave velocity.
Consistent with this identification, observe the linear dependence of 
$\overline{m}$ with $T$ in Fig.~\ref{mbar1} 
at $k_F =0.2$ over a wide temperature region.
This value of $k_F$ places the system quite close to the quantum-critical
point as the value of $\Delta$ is very small. 
At the quantum-critical point of the sigma model, it is predicted
that $\xi^{-1} = C_Q k_B T / \hbar c$ with $C_Q \approx 1$ a universal number.
The slope of $m$ versus $T$ at $k_F =0.2$
in Fig.~\ref{mbar1} is about $0.65$ - this matches with
the expected result if there is renormalization of spin-wave velocity
$c/c_0 \approx 1/0.65$. A renormalization of the spin-wave velocity
of order unity is to be expected, as the fermionic polarization
$\chi$ in (\ref{chival}) is not suppressed by a factor that vanishes
as $\delta \rightarrow 0$.
%
%

Differences between the quantum transition in the doped and undoped
antiferromagnet do however show up at the correction to scaling level.
The higher-order non-analytic terms in $\Sigma_f$ will 
have a form which is quite specific to the doped model. 
One such term can be obtained by analytically continuing to real frequencies
and computing $\mbox{Im} \Sigma_f$ at the critical point. We find
\begin{equation}
\mbox{Im} \Sigma_f ( q , \omega )
\sim q \omega^2 ~~~~\mbox{at the quantum-critical point}
\end{equation}

\subsubsection{Lifshitz point}
The existence of a direct CLRO to ILRO transition in Fig.~\ref{mbar2}
has a strong consequence for the phase diagram of the SS model.
As both phases can be transformed into their quantum-disordered
partners simply by increasing the value of $g$, we conclude that
there must be a point in the phase diagram where all the four
phases - CLRO, CQD, IRLO, and IQD - meet. Such a point is called
a Lifshitz point~\cite{horn}.
Lifshitz points have so far been studied primarily in the context of thermal
transitions in classical spin systems~\cite{selke}.
An important result is that such points can exist only above a lower
critical dimension determined as follows~\cite{grest}: a system in
$D$ dimensions, with incommensurate instabilities in $m$ of those
dimensions, has lower critical dimension $2 + m/2$. This result
appears to be in conflict with our results here for the doped
antiferromagnet. For we have
incommensuration in $m=2$ spatial dimensions, giving a lower critical
dimensionality equal to the spacetime dimension $D=3$. So how can a 
Lifshitz point exist ?

The answer to this apparent inconsistency lies in the form of
$\Sigma_f$. At the Lifshitz point we clearly have $\partial G^{-1} /
\partial q^2 |_{q=0,\omega=0 } = 0$. Thus the leading $q$ dependence
of $G^{-1}$ at small $q$ will come from higher-order terms in 
$\Sigma_f$. Let us assume that $G^{-1} \sim \omega^2
+ q^p$ at the Lifshitz point. Inserting this fully renormalized
$G$ in the result (\ref{Sigma}) for $\Sigma_f$ we find by power counting 
\begin{equation}
\Sigma_f ( q, 0) = \Sigma_f (0,0) + b_1 q^2 + b_3 q^{4-p/2} + \ldots
\end{equation}
Consistency now demands that $p=4-p/2$ which yields $p=8/3$.
This differs from the 
value $p=4$ use in classical spin systems~\cite{selke}. With this
modified form of $G$ we may repeat the calculation of
Grest and Sak~\cite{grest} and verify that spacetime dimension $D=3$ is above
the lower critical dimension which is $D=7/3$. 
Thus it is possible to have a Lifshitz point in $D=3$.

Finally we note that a point where CLRO, ILRO, CQD, and IQD phases
meet was also found in the large $N$, $Sp(N)$ theory of frustrated,
two-dimensional quantum Heisenberg antiferromagnets~\cite{Sach-Read2}.
The nature of this point appears to be quite
different from the Lifshitz point in the present theory. In
particular, the IQD phase of the $Sp(N)$ frustrated 
antiferromagnets~\cite{Sach-Read2} contains
deconfined, bosonic, spin-1/2 spinons, while here we have found
massive, triply-degenerate $n$ quanta. Perhaps related to this
difference is the fact that the large $N$ $Sp(N)$ theory~\cite{Sach-Read2} 
finds no softening
in the spinon spectrum at the CQD-IQD boundary. Instead, at the
boundary the parabolic
spinon spectrum splits into two parabola with minima at
incommensurate points; the curvature at the minima
of the parabola always remains finite.
Contrast this with the behavior of the $n$ spectrum found here:
the curvature at the minimum of the $n$ spectrum vanishes at the
CQD-IQD boundary leading to double minima at incommensurate points
in the IQD phase.

\section{CONCLUSIONS}
\label{conc}

The most notable features of our results on lightly-doped 
antiferromagnets are the quantum disordered
phases with a spin pseudo-gap. These phases posses fully-gapped,
triply-degenerate, spin 1 magnons, and gapless, spin-1/2,
charge $e$ fermions. Several other
investigators~\cite{nagaosa,bza,spn,sokol2,monien} have also recently
explored
models of the normal state of the lightly-doped cuprates which
have spin gaps/pseudo-gaps.
This interest in pseudo-gaps is of course motivated by numerous
experiments on the underdoped cuprates showing gap-like
features in the normal state.~\cite{walstedt,bucher,Hardy}
There are also interesting trends in the doping and temperature
dependence of uniform spin susceptibility of the
cuprates~\cite{millis}.

For completeness we review some of
the previous theoretical results, and point out the differences to
our results.
A number of the models~\cite{nagaosa} are related to
resonating-valence-bond type mean-field theories~\cite{bza}; there
is a BCS-like pairing of spin-1/2, neutral spinons in the normal
state at low-doping, leading to a gap-like feature in the spectrum
at finite temperature. However, unlike our results, this state
extrapolates to a true gap at $T=0$. An extension of these models
to a three-band, $CuO_2$ layer model~\cite{spn} did posses gapless,
spin-1/2, charge $e$ fermionic excitations on the oxygen sites.
However the above-gap spectrum in all of these models~\cite{nagaosa,spn}
consists of unbound spin-1/2, neutral fermions; in contrast, the
spin spectral weight above the gap of our model is dominated by
a spin-1, bosonic magnon.
Millis and Monien have attributed the gaps to interlayer 
couplings in the Yttrium based cuprates~\cite{monien}. The recent work of Sokol,
Pines and collaborators~\cite{sokol2} is perhaps closest in spirit to
ours, although their scenario for the mixing between the $n$ quanta
and the fermions appears to be somewhat different.

Also relevant to our result is the recent high temperature series analysis of
the CLRO to CQD transition in the $t-J$ model~\cite{sokol1}. This 
work provides some evidence in support of our result that $z=1$ at
this transition.

Finally, if our model is to provide a complete picture of the
cuprates, it should also explain the nature of the photo-emission
spectrum~\cite{photo}. For this one needs to understand better the
connection between the bare electron ejected in the photo-emission
and the $(n_{\ell}, \Phi_{va})$ fields. This problem is currently
under investigation.

\acknowledgments

The research was supported by NSF Grant No. DMR-9224290.
I am indebted to A. Chubukov for valuable
discussions on every aspect of this work, for his comments on
the manuscript,
and for our previous collaborations on related subjects~\cite{comment,CSY}.
Jinwu Ye participated in some of the initial stages of this
work.
I thank N. Read, and B. Shraiman for useful discussions,
and ITP Santa Barbara for hospitality while part of
this work was completed.

\begin{figure}
\caption{Feynman diagrams for the $n$ self-energy. The thick, full line 
is the $n$ propagator, the thin, full lines are the hole fermions, and the 
dashed line is the $\lambda$ propagator.}
\label{feyn}
\end{figure}
\begin{figure}
\caption{Values of $\overline{m}$ (defined in 
Eqn. (\protect{\ref{defmb}}); it is roughly the inverse correlation length)
as a function of $T$ for various doping levels specified by $k_F$.
The coupling constants were $g=3$, $\kappa = 1.5$, 
$(m_{\ell} m_{h})^{1/2} = 0.6$, $m_{\ell}/m_h = 0.1296$. 
The maxima in the structure factor are at $q = q_c (\pm 1, \pm
1)/\protect{\sqrt{2}}$ in our rotated co-ordinate system; this places them
at $(\pi \pm q_c , \pi)$ and $(\pi, \pi \pm q_c)$ in the conventional
Brillouin zone of the square lattice.
All parameters are measured
in the dimensionless units described in Section~\protect{\ref{calcs}}. The
small $T$ behavior of $\overline{m}$, and the value of $q_c$
identifies the nature of the ground state (CLRO, ILRO, CQD, IQD)
which is also noted.
}
\label{mbar1}
\end{figure}
\begin{figure}
\caption{The values of $m$ associated with the results for
$\overline{m}$ in Fig.~\protect{\ref{mbar1}}. For the commensurate
states $m=\overline{m}$.}
\label{m1}
\end{figure}
\begin{figure}
\caption{As in Fig.~\protect{\ref{mbar1}} 
but for $g=6$, $\kappa = 2$, 
$(m_{\ell} m_{h})^{1/2} = 0.6$, $m_{\ell}/m_h = 0.1296$.
}
\label{mbar2}
\end{figure}
\begin{figure}
\caption{The values of $m$ associated with the results for
$\overline{m}$ in Fig.~\protect{\ref{mbar2}}.}
\label{m2}
\end{figure}
\begin{figure}
\caption{
Contribution of the $n$ field to the structure factor,
$\tilde{S}(q)$, in the IQD phase. We use the coupling constants
of Fig.~\protect\ref{mbar1} at $k_F = 0.3$. The temperatures
are ({\em a\/}) $T=0.5$, ({\em b\/}) $T=0.2$, and ({\em c\/})
$T=0.02$. The jagged double-peaks in ({\em c\/}) are an artifact of
the plotting routine. 
}
\label{sfac}
\end{figure}

\begin{references}
\bibitem{square} The system does not have full rotational invariance,
only the symmetry of the square lattice: the notation $q^p$ is 
a schematic for a function of $q_x$, $q_y$ which is homogeneous
of degree $p$ and is invariant under the square lattice space group.
\end{references}
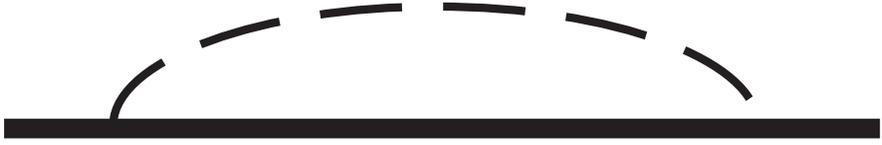
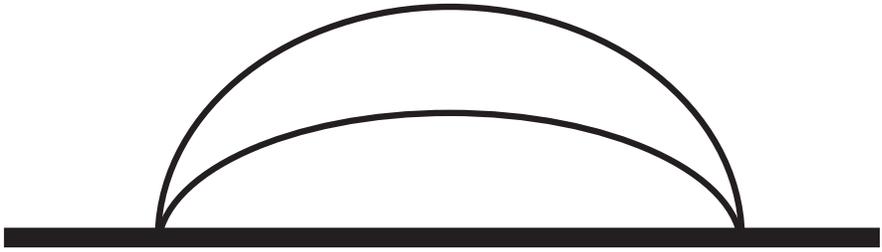
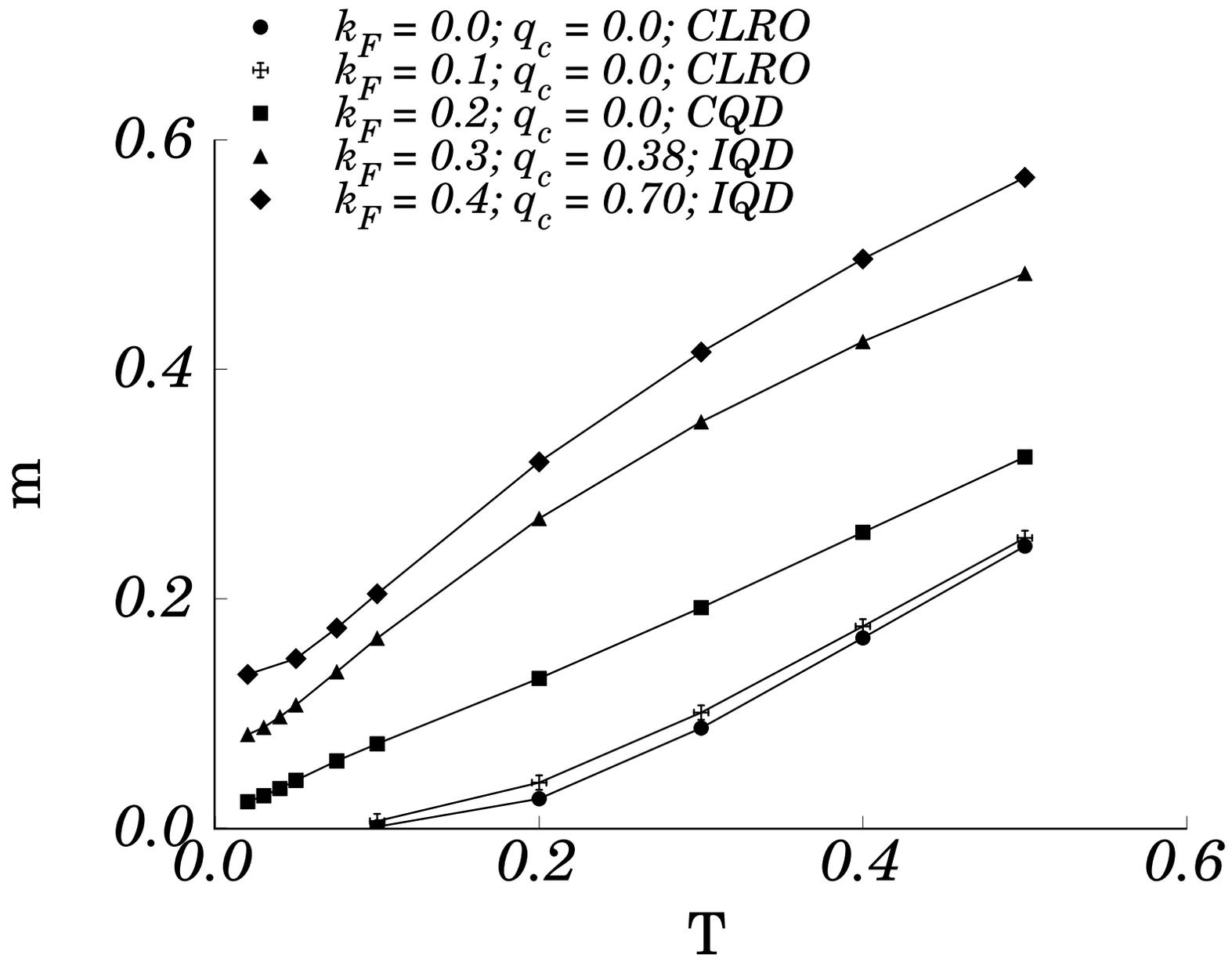
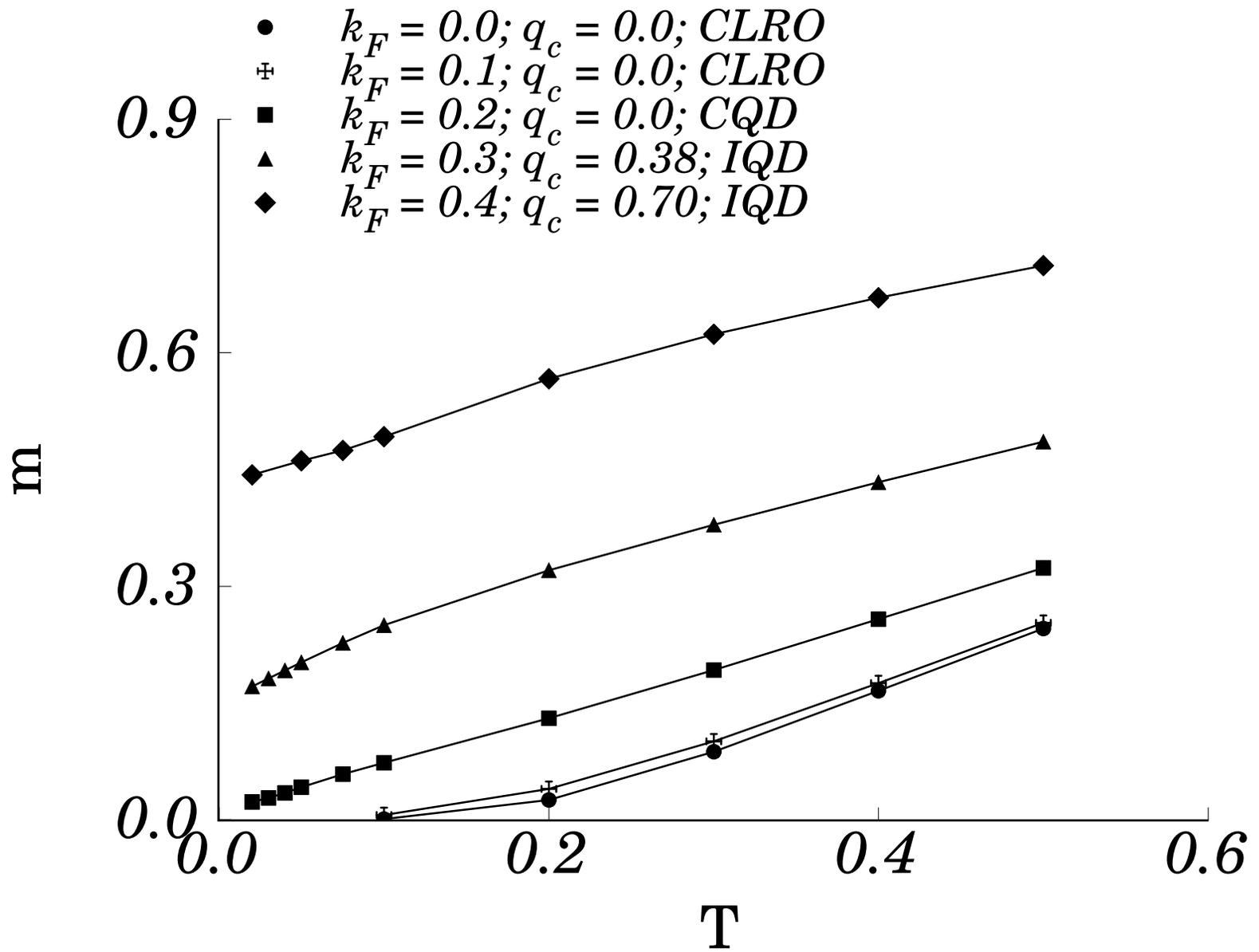
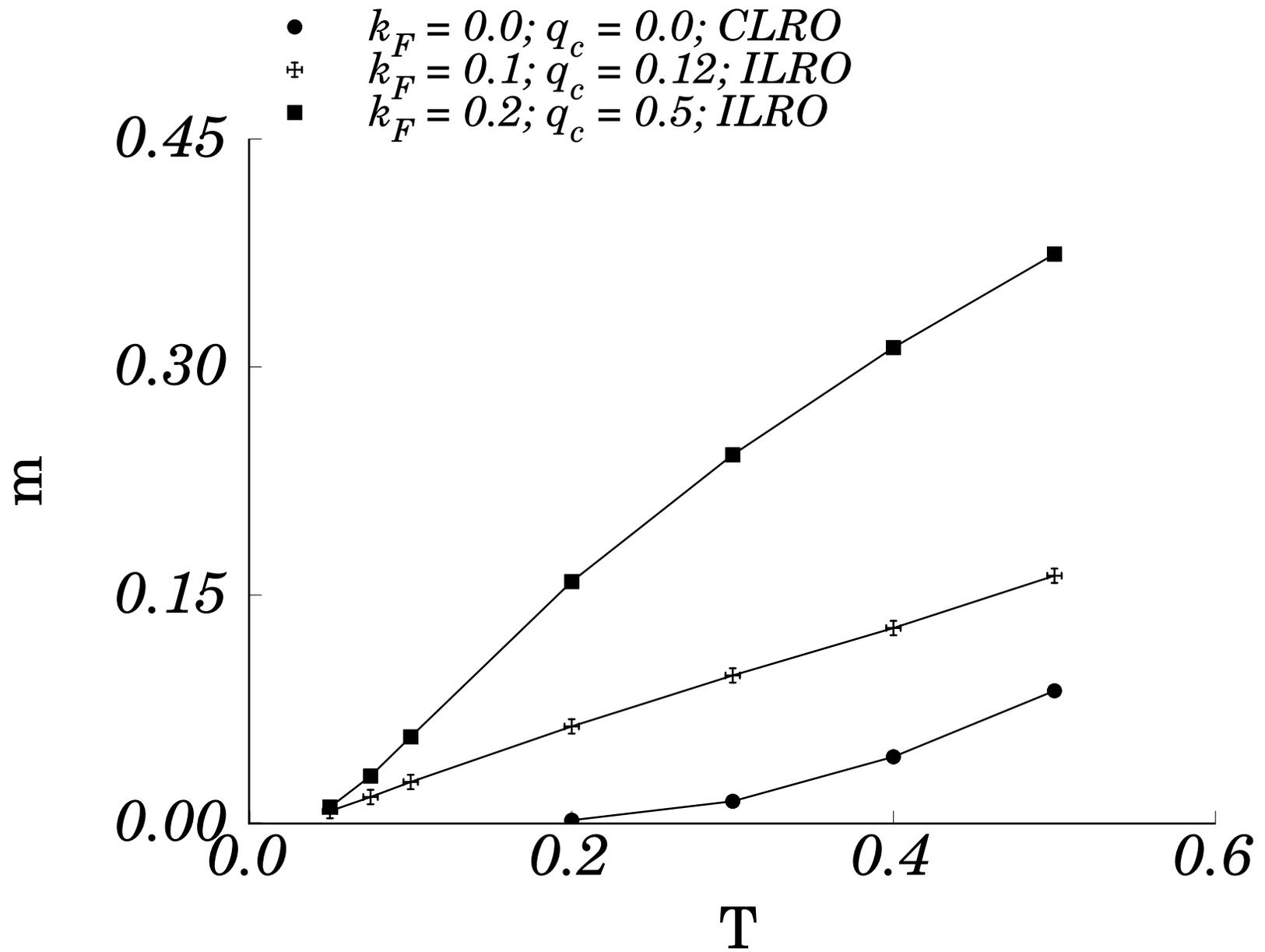
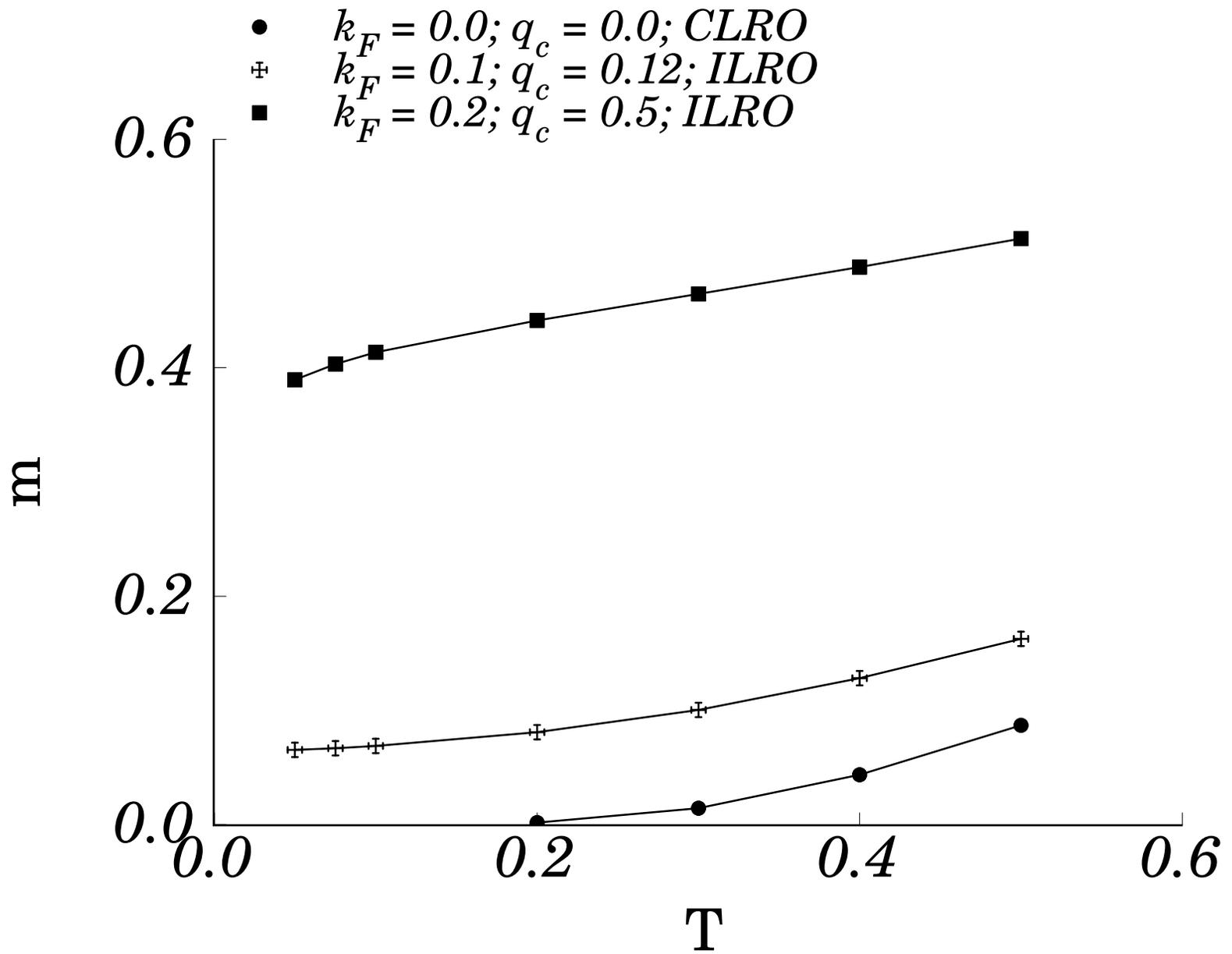
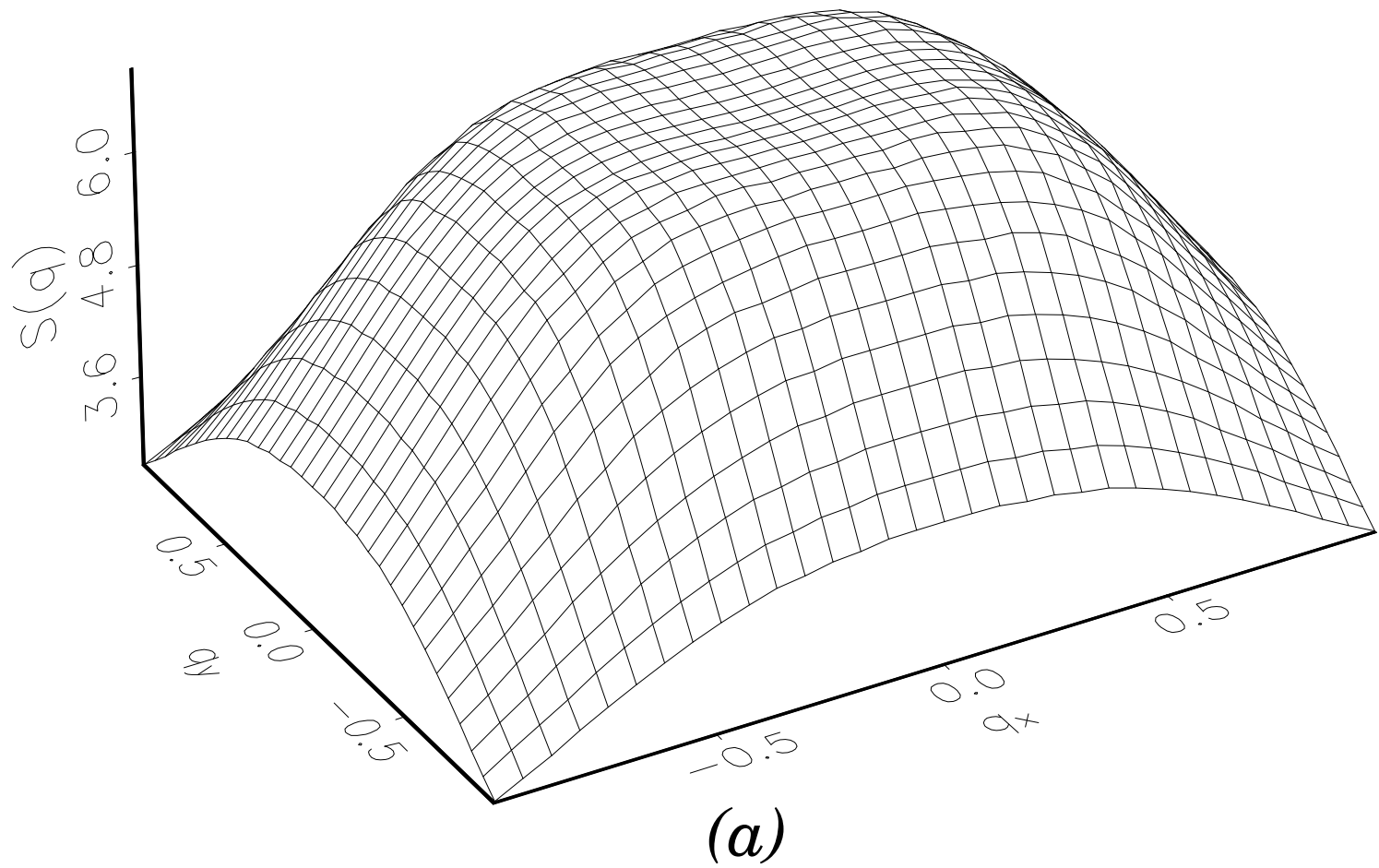
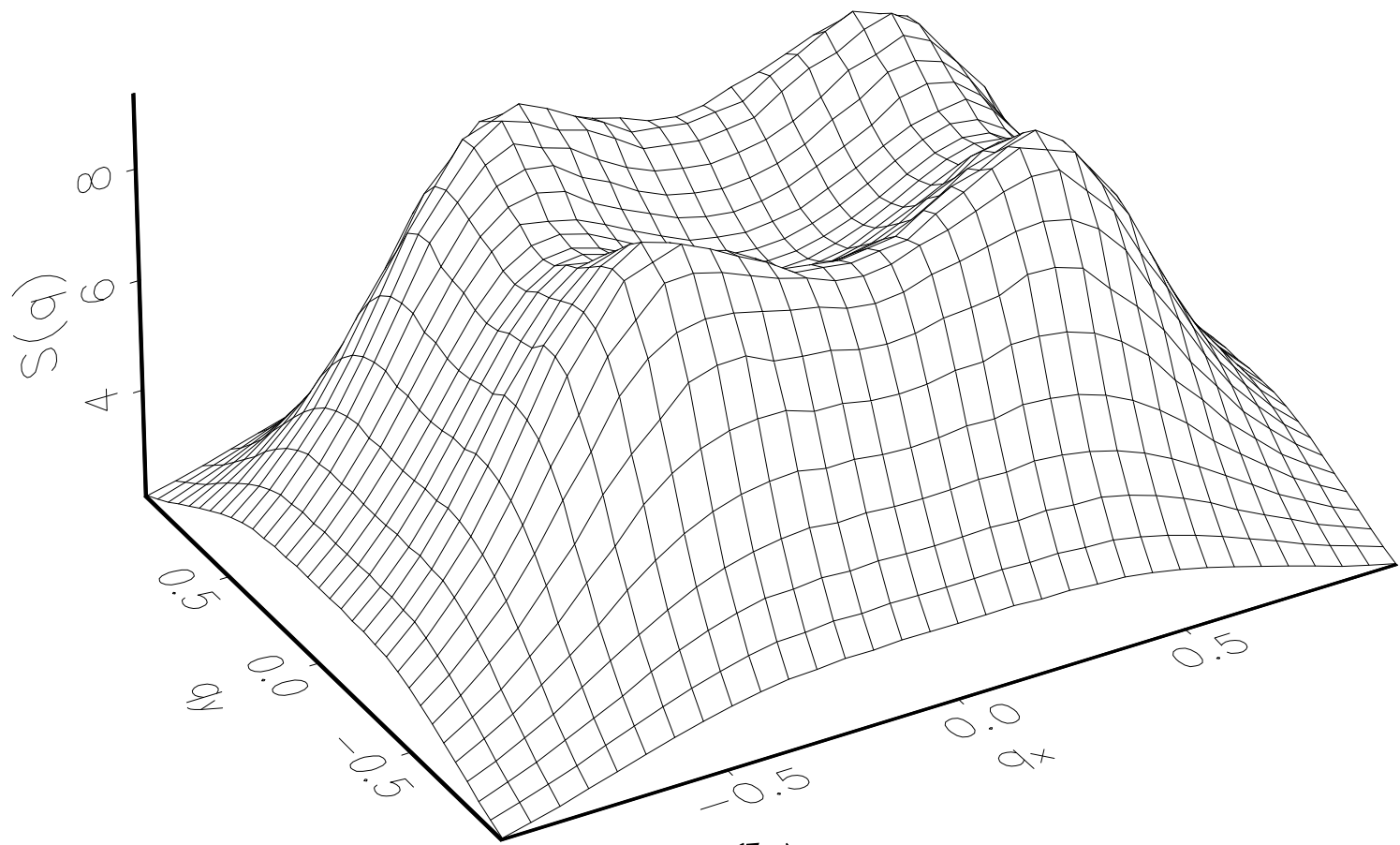
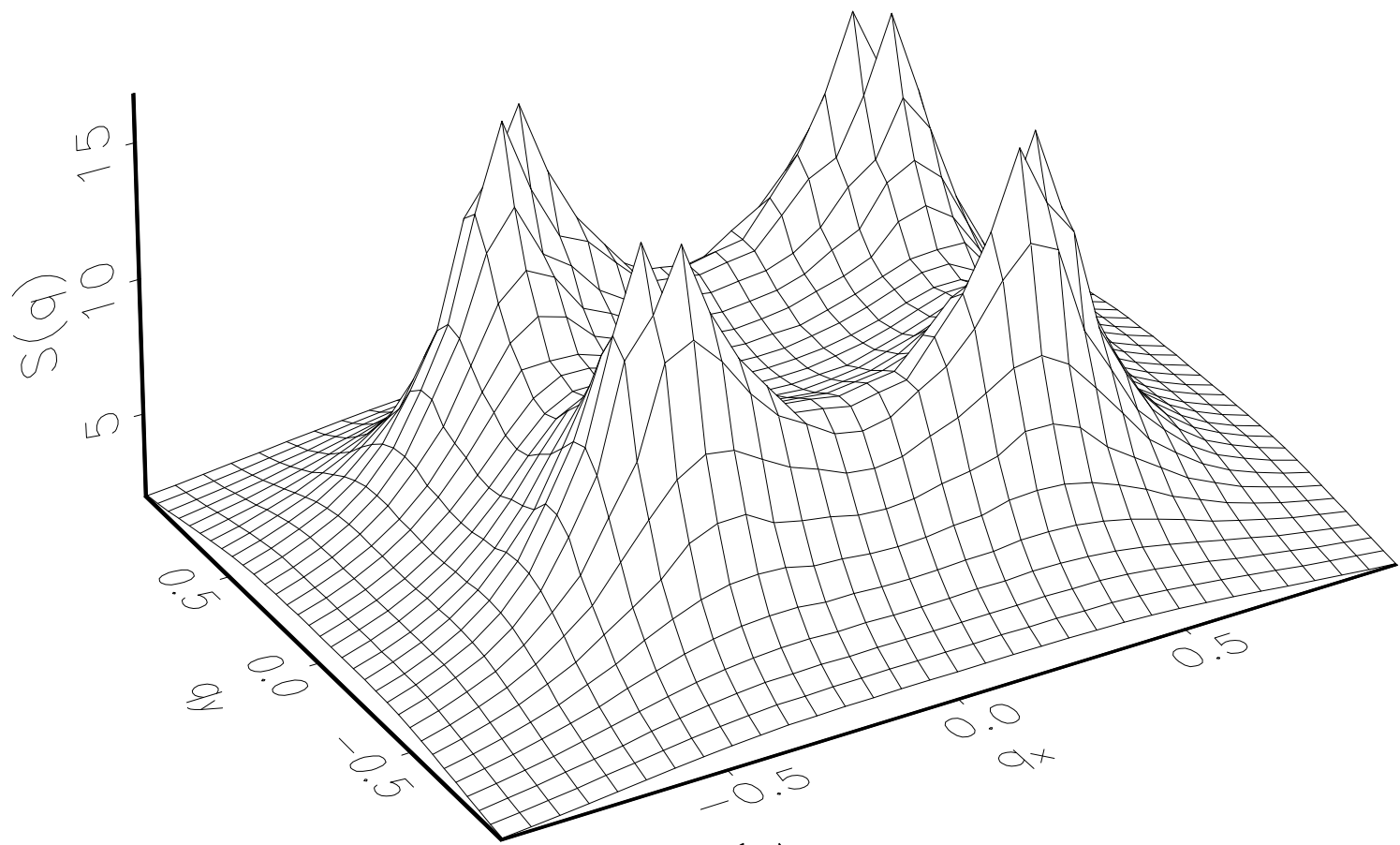
\end{document}